# NONPERTURBATIVE EVALUATION OF THE SHPALERON TRANSITION RATE

Talk given at NATO Advanced Research Workshop on Electroweak Physics and the Early Universe, Sintra, Portugal, 23-25 March 1994

A. Bochkarev

Department of Physics,
University of Pittsburgh,
Pittsburgh, PA 15260

*"You haven't measured it, you've only calculated it..."*
*Jan Smit.*

We review nonperturbative calculations of the rate of sphaleron transitions on the lattice in $1+1$-dimensional field theories and introduce a way to perform gauge-invariant Gibbs averages in classical non-Abelian Higgs theories.

Effective potential of the Chern-Simons variable $N$ in the Standard Model (SM) is bounded at infinity. For this reason in hot electroweak plasma one expects Brownian motion of the system between topologically distinct vacua labelled by integer vallues of $N$ :

$$\Delta(t) \;=\; \ll (N(t) - N(0))^2 \gg \;\to\; \Gamma\, t\;,\quad as\;\; t \to \infty, \tag{1}$$

where $\ll \cdot \gg$ stands for the Gibbs average. Diffusion rate $\Gamma$ is the rate of sphaleron transitions. The parameter $\Delta(t)$ at large times receives contributions only from the relevant fluctuations interpolating between different classical vacua, which makes it possible in practice to calculate $\Gamma$ numerically on the lattice. $\Gamma$ is a nonperturbative definition of the rate of anomalous baryon number nonconservation at high temperatures in the SM [1], which can be used in the semiclassical temperature domain and beyond it. At high temperatures one neglects quantum corrections and solves classical equations of motion to generate static solutions like kinks [2] or sphalerons [3].

The correlator (1) was first evaluated in $1+1$-dimensional Abelian Higgs model in [4] and the Brownian behaviour of the Chern-Simons number was found. Two methods to evalute $\Delta(t)$ are known so far. The first one is based on the microscopical classical *real-time* Langevin equation, derived in [4] in such a way that the white noise would respect gauge invariance. The Lagrangian of the model is:

$$\mathcal{L} \;=\; -\frac{1}{4\xi} F_{\mu\nu}^2 \;+\; |D_\mu \phi|^2 \;+\; (|\phi|^2 - 1/2)^2\;, \tag{2}$$

with $D_\mu = \partial - iA_\mu$. In this Abelian gauge theory one can resolve constraints in the Coulomb ($\partial_1 A_1 = 0$) gauge using polar coordinates $\phi = \rho e^{i\alpha}$. The only (gauge-invariant!) degree of freedom left of the gauge field after the constraints have been resolved is the Chern-Simons variable $N = 1/2\pi \int dx\, A_1$. The momentum $p_\alpha$ canonically conjugated to the gauge-depenent degree of freedom $\alpha$ is the the density of electric charge. The white noise cannot be introduced in the equation for $\dot{p}_\alpha$, because it would violate the conservation of the electric charge. Therefore the noise term must be introduced only to the gauge-invariant variables like $\rho$. The solution $N(\tau)$ of the resulting first-order Lagevin equation is obtained numerically and the diffusion parameter (1) is evaluated as

$$\Delta(t) = \frac{1}{\tau_o} \int_o^{\tau_o} d\tau \ (N(t+\tau) - N(\tau))^2, \quad as \ \tau_o \to \infty, \qquad (3)$$

In this way the diffusion (1) was observed [4] with the expected thermal activation behaviour of the diffusion rate $\Gamma \sim \exp(-E_{sp}/T)$ at $T < E_{sp}$, where $E_{sp} \simeq 1$ is the sphaleron mass.

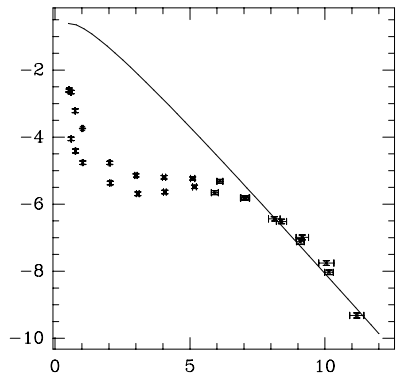

Figure 1: $\ln(\Gamma)$ vs $\beta$ for lattice spacings $a = (0.16, 0.32)$ in the Abelian Higgs model [5].

The second method is to calculate the Gibbs average (1) directly according to the definition of the classical Green function:

$$\Delta(t) = \mathcal{Z}^{-1} \int \mathcal{D}p \, \mathcal{D}\phi \ \exp\left[-H(p,\phi)/T\right] \Delta_{p\phi}(t) \qquad (4)$$

where $\Delta_{p\phi}(t)$ is evaluated on the microcanonical trajectories $p(t), \phi(t)$ - solutions of the Hamiltonian equations of motion with the initial conditions $p(t) = p$, $\phi(t) = \phi$. The Metropolis algorithm generating initial configurations must respect the Gauss law. For this purpose the gauge-invariant Langevin equation of [4] was used in [5] to generate the initial $\{\phi, p\}$. The temperature dependence $\Gamma(T)$ of these canonical measurements of the diffusion rate [5] is shown on Fig.1. At semiclassical temperatures $\beta \equiv T^{-1} > 7$ the data agree well with the analytical predictions of [6], while at high temperatures $\beta < 3$ the rate exhibits the expected $T^2$-law. In [7] it was found numerically advantageous to use $A_o = 0$ - gauge in Cartezian coordinates during the Hamiltonian evolution.

The gauge-invariant Langevin equation of [4] was introduced in the Abelian theory, where one can resolve the gauge constraints explicitly. To generalize it to the non-Abelian case we suggest to use the unitary gauge. In SU(2)-gauge theory with a scalar doublet $\varphi$ the unitary gauge implies $\varphi = \sqrt{\varphi^\dagger \varphi} \equiv \rho$, and the Hamiltonian

$$\mathcal{H} = \frac{1}{2\xi}E^2 + \frac{1}{2\xi}H^2 + \frac{1}{4}p^2 + (\nabla \rho)^2 + \rho^2 A_i^2 + V(\rho) + \frac{1}{4\xi^2\rho^2}(D_i E_i)^2 \qquad (5)$$

can be expressed entirely in terms of canonically conjugated variables $(\rho, p)$ and $(A_i^a, E_i^a)$. The last term (Coulomb energy $\rho^2 A_o^2$) takes into account the Gauss law $D_i E_i = -2\rho^2 A_o$ to get rid of $A_o$. The Gauss law is automatically satisfied in, say, Metropolis samplings with the Hamiltonian (5). In this way the gauge-invariant Gibbs distribution in classical non-Abelian gauge theories can be performed by Monte-Carlo lattice simulations. Field configurations generated in the unitary gauge do not contain sphalerons. The latter appear during the canonical evolution according to the equations of motion, which are gauge-invariant.

Langevin and canonical times are two alternative definitions of physical time in the processes at $T \neq 0$. To find a relation between them the sine-Gordon model was studied [8]:

$$\mathcal{H} = \int dx \left\{ \frac{1}{2}(\dot{\phi})^2 + \frac{1}{2}(\triangle \phi)^2 + 1 - cos(\phi) \right\} \qquad (6)$$

The diffusing variable now is the zero mode $\bar{\phi} = \frac{1}{2\pi L} \int dx\, \phi(x)$. The model has many degenerate classical vacua and stable kink solutions interpolating between them. At $T \neq 0$ the Brownian motion of $\bar{\phi}$ has been observed. At semiclassical temperatures it is due to creation of kink-antikink pairs, so the thermal activation-type of the temperature dependence $\Gamma \sim exp[-E_K/T]$ has been seen in both Langevin and direct Gibbs measurements.

The complete second-order Langevin equation has friction coefficient $\gamma$ as an input parameter:

$$\ddot{\phi}_x + \frac{\partial \mathcal{H}}{\partial \phi_x} = -\gamma \dot{\phi}_x + \eta_x(t), \qquad (7)$$

with the white noise normalized as $<\eta_x(t)\eta_x(0)> = 2T\gamma\delta(x)\delta(t)$. Eq. (7) interpolates between the Hamiltonian equation of motion for $\gamma \to 0$ and the first-order Langevin equation (no $\ddot{\phi}_x$-term in (7)), valid at large $\gamma$ or $t$. The absolute value of the rate (the preexponential factor) depends on $\gamma$. For moderate to large friction the simulations of (7) show (Fig.2): $\Gamma(\gamma) \sim \gamma^{-1}$; which would be a prediction of the first-order Langevin equation. That is because $\Gamma$ is extracted from the correlator at large times $t$, for which the first-order Langevin equation is usually derived.

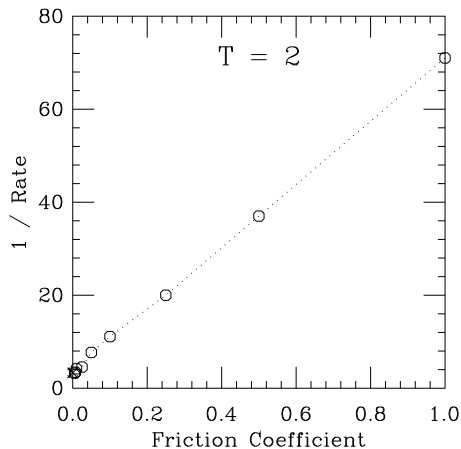

Figure 2: $\Gamma^{-1}$ vs friction coefficient $\gamma$ at $T = 2$ in the sine-Gordon model [8].

In the limit $\gamma \to 0$ the rate remains nonzero and has been found [8] to coincide exactly with the corresponding independent direct Gibbs measurement according to (4). These data demonstrate a fundamental conclusion that canonical measurements can be obtained as the $\gamma \to 0$-limit of the second-order Langevin simulations.

In the high-temperature domain $T > E_K$ at moderate to large friction the second-order Langevin simulations [8] show: $\Gamma = 2T/\gamma$, which is obviously the prediction of the *free* first-order Langevin equation. These simulations are insensitive to the dynamics of the system, because at high temperatures the noise-term $\eta_x(t)$ entirely dominates over the regular Hamiltonian force.

The high-temperature domain of canonical measurements was studied in [9]. It corresponds to the second-order Langevin simulations with asymptotically small $\gamma$, so that the noise-term does not dominate over the regular force at any fixed $T$. This set up represents a large isolated system like the early Universe with no external dumping. In addition to the expected temperature behaviour $\Gamma = T^2$ a strong lattice spacing dependence of the diffusion rate was discovered (Fig.3): $\Gamma \sim a^{-2}$. This is in contrast with the moderate to large friction simulations which do not exhibit any lattice-spacing dependence for $a \leq 1$ [8]. The rate, however, remains remarkably constant for the fixed ratio $T/a$ even as one goes into the semiclassical temperature domain [9]. Presumably, damping with the friction $\gamma$ screens the high-frequency ultraviolet modes $\omega \sim a^{-1} \geq \gamma$, which are essential beyond the semiclassical domain of temperatures. The lattice spacing dependence of the diffusion rate may be different between the sine-Gordon and Abelian Higgs models, because the dimensionality of the rate is different. The latter is fixed by the requirement that the diffusing variable is dimensionless, since it has integer dimensionless values in the classical vacua. Study of the lattice spacing dependence of the diffusion rate requires rather long simulations. As $a$ decreases, the number of degrees of freedom grows, and it takes more and more time for the system to thermalize itself in the course of the canonical evolution, which always takes place before the Brownian motion sets in. This study is, however, important particularily in higher space-time dimensions, since it is a connection between the simulated classical system and the underlying quantum field theory at finite temperature [2], [10].

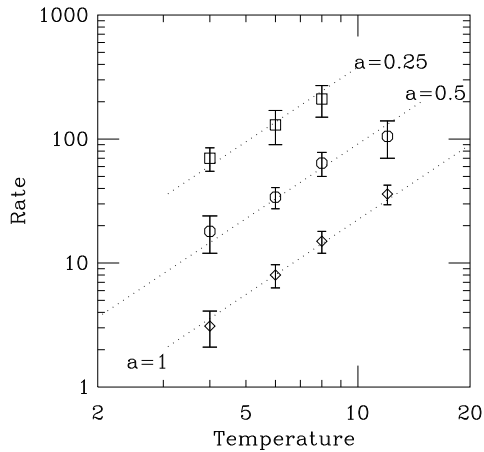

Figure 3: $\Gamma(T)$ for $a = (1, 0.5, 0.25)$ at high $T$ in the sine-Gordon model [9].

This work was supported by the U.S. NSF grant PHY90-24764 and the grant of the Pittsburgh Supercomputer Center.